# Systematically Examining Reproducibility: A Case Study for High Throughput Sequencing using the PRIMAD Model and BioCompute Object

Meznah Aloqalaa[1], Stian Soiland-Reyes[1] & Carole Goble[1]

## Abstract

The reproducibility of computational pipelines is an expectation in biomedical science, particularly in critical domains like human health. In this context, reporting next-generation genome sequencing methods used in precision medicine spurred the development of the IEEE 2791-2020 standard for Bioinformatics Analyses Generated by High-Throughput Sequencing (HTS), known as the BioCompute Object (BCO). Championed by the USA's Food and Drug Administration, the BCO is a pragmatic framework for documenting pipelines; however, it has not been systematically assessed for its reproducibility claims.

This study uses the PRIMAD model, a conceptual framework for describing computational experiments for reproducibility purposes, to systematically review the BCO for depth and coverage. A meticulous mapping of BCO and PRIMAD elements onto a published BCO use case reveals potential omissions and necessary extensions within both frameworks. This underscores the significance of systematically validating claims of reproducibility for published digital objects, thereby enhancing the reliability of scientific research in bioscience and related disciplines.

[1] The University of Manchester, Department of Computer Science, Manchester, M13 9PL, UK

# 1. Introduction

Reproducibility is fundamental in science as a means of validating and strengthening studies, giving researchers confidence in their results and enabling them to build on these findings (Innovation and Technology Committee, 2023). This principle has received increased attention in the scientific field in recent decades (Ioannidis, 2005; Donoho et al., 2009; Pashler et al., 2012), including concerns voiced as a 'reproducibility crisis' (Baker, 2016). Numerous initiatives launched in scholarly communities across different domains have sought to highlight the state of the art and best practices in different communities including psychology (Open Science Collaboration, 2015), biology (Errington et al., 2021), computer science (Collberg and Proebsting, 2016), social science (Hardwicke et al., 2020) and artificial intelligence(AI) (Hutson, 2018; Gundersen and Kjensmo, 2018). As computational methods have become more prevalent in research (Freire et al., 2012) so the reproducibility of analysis workflows, simulations, digital twins, and AI predictions has come to the foreground. Although reproducibility is thought to be expeditable by computational approaches, challenges arise even when automation covers all or most of the steps in research experiments (Crook et al., 2013). For example, in a recent effort to reproduce crash-free executions for 101 scientific workflows resulted in only a 28% success rate (Grayson et al., 2023). This is despite the development of many approaches for enhancing computational reproducibility, for example: research outputs repository platforms (e.g., Zenodo, Figshare); research packaging approaches (e.g., ReproZip, Whole Tale, RO-Crate); workflow management systems (e.g., Snakemake, Nextflow); containers (e.g., Docker, Singularity); computational notebooks (e.g., Jupyter); and version control systems (e.g., Git, GitHub).

Efforts have been made to define reproducibility and compare it with the concepts of replicability, robustness, generalisability and reusability, as well as the overarching issue of transparency (Claerbout and Karrenbach, 1992; Stodden, 2014; Goodman et al., 2016; NASEM, 2019; Community TTW, 2019; ACM Artifact Review and Badging, 2020; Gundersen, 2021;). Initiatives to incentivise adherence include measures such as the reproducibility badging practices of research publishers (ACM, IEEE, Elsevier, and Springer Nature (NISO, 2021)). The tendency towards "R" wordplay (reproducible, repeatable, reusable, retrievable, reprocessable, repurposable, reliable, recoverable, etc.) can become overwhelming (Roure, 2014). The terminology of *reproducibility* and *replicability* unfortunately is not consistent across disciplines (Plesser, 2017), we here use the definitions of NASEM (2019) with respect to computation:

- Reproducibility: "obtain consistent computational results using the same input data, computational steps, methods, code, and conditions of analysis";
- Replicability: "obtain consistent results across studies aimed at answering the same scientific question, each of which has obtained its own data";
- Generalizability: "the extent that results of a study apply in other contexts or populations that differ from the original one".

## 1.1 Unpacking Reproducibility: The PRIMAD Framework

The NASEM definitions hint at fixity; that is, there are properties that remain the same, and there are those that are allowed or expected to be disrupted when results are compared. Such fixity is explicitly addressed in the PRIMAD framework (Freire et al., 2016, pp. 128–132) through systematic descriptions of computational experiments. The acronym 'PRIMAD' (Table[1]) represents the prime variables that can change (or be fixed) during the reproduction of a computational study (Freire et al., 2016, pp. 128–132):

- **P: Platform:** the execution environment and computational context. This dimension represents the different specifications of the execution environment in the study. It includes hardware, software, compilers, storage, and/or different-purpose platforms (such as management and dissemination).
- **R: Research Objectives** or goals of the study.
- **I: Implementation:** the computational code or source code.
- **M: Methods:** the algorithm, pseudocode and/or the methodology that fulfilment the study goals.
- **A: Actor:** the people who contribute to the study.
- **D: Data:** input data and parameter values of the source code.

Computationally-based reproducibility studies are labelled differently in PRIMAD. In Table[1], Computational Reproducibility Labels are associated with their gains and the way of achieving them. What dimensions will change, change accordingly, or will not change, as follows:

- **Repeat:** nothing is changed;
- **Relocate/Port:** the software platform changes, but the encoding of the algorithm itself remains the same;
- **Reuse/Repurpose:** the experimental set-up is reused or repurposed for a different question, assuming it to be trustworthy;
- **Re-code/Reinterpret:** the algorithm/pseudocode remains the same but is re-implemented. The software platform may change;
- **Ratify/Validate:** Changing the method allows the validation of the correctness of a hypothesis using a different methodological approach. This change will incur a change in the parameters, implementation, and possibly also the execution environment (platform). Additionally, a change in data could happen because the new methodology could require a different type of dataset.
- **Review/Independent Verify:** the actor is different to the instigator for independent verification;
- **Resample/Generalize:** the computational aspects remain the same, but they are applied to different data, which could imply changes in parameters too;
- **Reparameterization/Recalibration:** a specialisation of generalise (a specialisation of robustness), where the parameters are systematically changed (a "sweep" across the parameters) for code reruns over the same data; or an algorithm optimization where the parameter is adjusted to enhance the algorithm's performance;

In Table [1], the labels "Reuse/Repurpose" and "Review/Independent Verify" focus on changing the Research Objective and Actor dimensions consecutively without altering the core structure and setting of the experiment, which includes Method, Implementation, and Data (Row and Parameter) dimensions. Consequently, this makes these labels orthogonal to the remaining labels.

| Computational Reproducibility Label | P Platform | R Research Objective | I Implementation | M Method | A Actor | D - Raw data | D - Parameters | Reproducibility Gain |
|---|---|---|---|---|---|---|---|---|
| Repeat | - | - | - | - | - | - | - | Consistency Across Trials, Determinism |
| Port, Relocate | x | - | - | - | - | - | - | Cross-Platform Compatibility (Portability), Minimal Dependency (Flexibility) |
| Re-use/Re-purpose | - | x | - | - | - | - | - | Cross-Disciplinary Application (Apply code in different sitting), Resource Efficiency |
| Re-code, Reinterpret | (X) | - | x | - | - | - | - | Improved Code Quality, Correctness of Implementation, Expand Adoption, Enhanced Efficiency, flexibility |
| Ratify, Validate | (X) | - | (X) | x | - | (X) | (X) | Hypothesis Correctness, Validation via a Different Approach, Findings Robustness |
| Review, Independent Verify | - | - | - | - | x | - | - | Enhanced Transparency (Sufficiency information), Independent Verification |
| Resample, Generalize | - | - | - | - | - | x | (X) | Applicability across Different Settings |
| Reparameterization, Recalibration, Parameter Sweep, | - | - | - | - | - | - | x | Robustness, Adaptation to other Conditions (Sensitivity), Parameters Optimization |

Table 1: The PRIMAD model for reproducibility in a computational experiment; each reproducibility label is associated with its gain and the way of achieving it. What dimensions will change X, change accordingly (X), or will not change - . (adapted from Freire et al., 2016, pp. 128–132). The original table has been updated with new gains and labels, which are highlighted using underlined text to distinguish the additions clearly.

The NASEM (2019) specifications for reproducibility, replicability and generalizability could mapped to the computational reproducibility labels' of PRIMAD with harmonious gains as follows:

- **Reproducibility (NASEM):** Repeat (PRIMAD): Nothing Change: Result Consistency
- **Replicability (NASEM):** Validate (PRIMAD): The main change in Method, in addition to consequence changes in other dimensions: Findings Robustness.
- **Generalizability (NASEM):** Re-use (PRIMAD): Change in Objective/Purpose: Cross-Disciplinary Application.

PRIMAD (Freire et al., 2016, pp. 128–132) also introduces two further dimensions:

- **Consistency:** the success or failure of a reproducibility study evaluated with respect to whether results are consistent with the previous ones rather than identical.
- **Transparency:** the ability to look into all necessary components to be able to understand the path from the hypothesis to the results. The degree of transparency should be used as a measure for the degree of inspection possible on codes and the availability of data.

Consistency and transparency represent the non-direct dimensions of the PRIMAD model, as they are not explicitly mentioned in the model acronym, unlike the direct dimensions: **P:**Platform, **R:**Research Object, **I:**Implementation, **M:**Method, **A:**Actor, and **D:**Data.

## 1.2 Metadata Research Object for Reproducibility Reporting

Computational pipelines have been increasingly adopted in biomedical science, including critical domains like human health. Next-generation sequencing (NGS) analysis methods used in precision medicine (Alterovitz et al., 2018) have spurred the development of the IEEE 2791-2020 standard for Bioinformatics Analyses Generated by High-Throughput Sequencing (HTS) (Alterovitz et al., 2018) (Simonyan et al., 2017) (Mazumder et al., 2020). Known as the BioCompute Object (BCO), the framework is specifically tailored to the demands of reproducibility in regulatory settings, notably the evaluation, validation, and verification of bioinformatics pipelines for the US Food and Drug Administration (FDA) (King et al., 2022), by standardising the reporting of genomic data information for exchange between stakeholders (Alterovitz et al., 2018) (Simonyan et al., 2017). The framework recommends and combines multiple standards for the comprehensive documentation of NGS data and analyses (Alterovitz et al., 2018). These standards include Fast Healthcare Interoperability Resources, the Global Alliance for Genomics and Health, Common Workflow Language (Crusoe et al., 2022), FAIR data principles (Findability, Accessibility, Interoperability, and Reusability) (Wilkinson et al., 2016), and Research Objects (Bechhofer et al., 2013).

Research Objects represent semantic encapsulation mechanisms for the research products used and produced during the research life cycle. The data, software, computational and manual processes, and documentation that are necessary for exchange, reuse, and reproducibility are encapsulated, described, and interlinked using metadata. The BCO packages data and metadata tailored to the generation, computing, and sharing of genomic pipelines. Other examples of research object frameworks include the FAIR Digital Object (De Smedt et al., 2020) and the RO-Crate (Soiland-Reyes et al., 2022), which has also developed profiles for the provenance of workflow execution pipelines (Leo et al., 2024).

The BCO is a pragmatic framework developed by the life science community. The BCO framework has three top-level fields and eight domains, as shown in Table[2]. The three fields – a unique identifier, a reference to its schema version, and a lightweight cryptographic hash value - are required attributes meant to aid the detection of unintended BCO alterations. The domains Usability, Provenance, Execution, Description, and Input/Output (IO) are required, whereas Parametric, Extension, and Error are optional (Table [2]). Usability domain is free text, the rest are a set of structured fields. However, some fields in the compulsory domains are optional.

| Mandatory | BioCompute Object (BCO) | Optional |
|---|---|---|
| **The three top-level fields** | | |
| Unique identifier | Schema version's reference | Lightweight cryptographic hash value |
| **Usability Domain:** free text used terminologies that are used in the BCO name, keywords and external references to describe computational workflow in the following sections: the need, method, results, and how the result can be used/interpreted for the scientific use case | **Provenance Domain:** history, version, contributors and review status of the BCO | |
| | **Input/output Domain:** input and output files used by the computational workflow for execution. | |
| **Description Domain:** pipeline steps, input and output file relations, and external references for review. | **Execution Domain:** execution of a BCO in terms of scripts and the information required to run a pipeline. | |
| **Error Domain:** specifies the accepted error range in an empirical and algorithmic manner. | **Parametric Domain:** a set of parameters used to customise the pipeline | |
| **Extension Domain:** the part of the workflow used to encompass additional structures outside the BCO schema | | |

Table 2: The BCO framework overview. The top-level fields and its eight domains.

Bioinformatics platforms such as Seven Bridges, DNAnexus, Galaxy, the High-Performance Integrated Virtual Environment (HIVE), and the Cancer Genomics Cloud (CGC) have developed support for BCOs, and a suite of tools (Xiao et al., 2020) (Patel et al., 2021), including a BCO portal and editor, have been built around the standard.

## 1.3 Examining the Reproducibility of BioCompute Objects

Despite the promotion of BCO's as a reproducibility standard for regulatory analysis, claims regarding the suitability of the BCO framework for reproducibility reporting have yet to be systematically examined. The BCO can be thought of as an implementation and interpretation of the reproducibility approach promoted by the PRIMAD model, which serves as a roadmap for the reproducibility of computational experiments. By cross-mapping the PRIMAD and BCO frameworks, we can systematically test the BCO's claims for reproducibility reporting and PRIMAD's robustness as a useful mechanism for thinking about reproducibility.

In the mapping between both frameworks, we used a published BCO as a use case. The findings highlight the value of PRIMAD as a framework for systematically validating claims of reproducibility for published digital objects, which we elaborate on in the discussion and conclusions.

## 2. Methods

To initiate the systematic mapping between the BCO and PRIMAD frameworks, we used a published BCO 'Regulatory BioCompute Object for Hepatitis C Virus Resistance Analysis' (BCO_022531/2.2) (King et al., 2022) as a use case. This use case represents the workflow that underlies the analysis of resistance to 3D treatment for hepatitis C genotype 1 and illustrates the failure of a treatment combination for the disease. The treatment combines 'a protease inhibitor (paritaprevir), an NS5A inhibitor (ombitasvir), and a polymerase inhibitor (dasabuvir)'. Treatment failed in 18 out of a sample of 100 patients in a simulated clinical trial for resistance against 3D treatment. The regulatory (BCO_022531/2.2) represents an NGS pipeline on the HIVE analysis platform designed specifically for the genomics and bioinformatics fields (Simonyan et al., 2016). The use case pipeline involves two primary steps ('HIVE-Hexagon' and 'Sequence Profiling Engine'), each reported through specific BCO domains, particularly within the Description, Execution, Parametric, and Error domains.

The systematic mapping between BCO and PRIMAD analyses used an examination phase followed by mapping phases as the following steps:

**Step 1:** A comprehensive examination of BCO_022531/2.2 explored the BCO specification;

**Step 2:** A general mapping of the PRIMAD model to the BCO framework using just the concepts (their specifications);

**Step 3:** A mapping of the BCO use case (BCO_022531/2.2) domains' fields to/from the PRIMAD model dimensions;

For the purpose of this paper, the notation PRIMAD.Dimension refers to the specific dimension of the PRIMAD model, PRIMAD.Dimension (sub) refers a sub-dimension, meanwhile BCO.Domain refers to the specific domain of the BioCompute Object, and *fields* or *subfields* as an italic text denotes the field and subfields of the BCO Domains. For consistency with tables, we will use an English shortform of the BCO domains, while the fields match the corresponding JSON keys.

# 3. Results

The results of the systematic mapping steps are provided in the Data Record section. Sections 3.1 will discuss the results of the comprehensive examination of the usee case:BCO_022531/2.2 (Step1 ) while section 3.2 will discuss the results of the general mapping and the mapping using BCO_022531/2.2 (Step 2 and 3).

## 3.1 Comprehensive Examination of BCO_022531/2.2

The examination is presented in a tabular representation of BCO_022531/2.2, including domains and their attributes, viewable from the BCO Portal. Our analysis was based on the BCO specification version 22.01, released on February 9, 2022, encompassing the BioCompute object user guide and the IEEE 2791-2020 schemas. Information on domains was extracted from the BCO's JSON. BCO_022531/2.2 generally adheres to the requirements of the BCO specification; however, in practice, we uncovered a number of issues:

- **Insufficient licence.** The compulsory field *license* is filled in with the URL https://opensource.org/licenses/MIT, but the link directs one to a license template. Information on the year and the copyright holder is not explicitly stated in this BCO.
- **Missing or obfuscated field values.** Files have no descriptive names that reflect their primary functionality (e.g., P0641M00002_S2_L001_R2_001 and P0641M00002_S2_L001_R1_001).

Such file naming conventions are common in laboratory or instrument outputs, but their meanings are difficult to decipher outside of a particular setting.

- **Access to platforms.** Accessing the use case workflow platform (HIVE) required time and consent from authors to activate the account and open the referenced files. Therefore, we noticed the files and script links have empty pages, and we had to inquire to determine whether access to them requires authorisation. File retrieval or re-computation on the basis of the BCO cannot be carried out independently of its platform of origin.

## 3.2 Mapping the PRIMAD Dimensions and BCO Domains

Table [3] summarises the mapping between PRIMAD dimensions and BCO domains (steps 2 and 3). The general mapping (step 2) and the mapping within the BCO use case (BCO_022531/2.2) (step 3), as expected, are not isomorphic. We apply (BCO_022531/2.2) to highlight the objectives of certain BCO fields with respect to the PRIMAD dimensions.

The mapping result is presented with respect to the PRIMAD model dimensions as follows:

**PRIMAD.Platform:** HIVE is the leading platform in the BCO.Description that holds the BCO files, and is also the platform that runs the script under BCO.Execution.

**PRIMAD.ResearchObjective** maps only to the *name* field in BCO.Provenance.

**PRIMAD.Implementation** maps to multiple fields in BCO.Description and BCO.Execution.

**PRIMAD.Method** maps to BCO.Usability but not BCO.Description because the pipeline steps are platform-dependent (HIVE).

**PRIMAD.Actor** maps to BCO.Provenance *contributor* field and its subfields: {*name, contribution, affiliation, email* and *ORCID*}.

**PRIMAD.Data (Raw)** includes data input (raw only) and maps to both BCO.IO and the input list fields in BCO.Description. There are three input files in *input subdomain* of BCO.IO and four in *input list* of BCO.Description (one for step1 and three for step2 of the pipeline).

**PRIMAD.Data (Parameter),** data parameters represent parameters that are used to measure an algorithm's robustness and sensitivity as the initial parameters for an experiment. In BCO.Parametric, a parameter list is used to customise a pipeline to affect results, recognising that humans "tweak" the pipeline. In the use case pipeline, there are 24 parameters for step 1 and 32 parameters for step 2.

| BioCompute Object | | | PRIMAD Model | | |
|---|---|---|---|---|---|
| Domain | Fields | Sub Fields | Dimension | | |
| Usability | | Free Text | Research Objective | Consistency | Transparency |
| | | | Method | | |
| Provenance | Review * | Name | | | |
| | | Version | | | |
| | | status | | | |
| | | reviewer comment | | | |
| | | date | | | |
| | | reviewer | | | |
| | | reviewer details | | | |
| | Derived From * | | | | |
| | Obsolescence Time * | | | | |
| | Embargo Period * | | | | |
| | Created Time | | | | |
| | Modification Time | | | | |
| | Contributors | Name | Actor | | |
| | | Contribution | | | |
| | | other details | | | |
| | Licence | | | | |
| Description | Keywords | | | | |
| | External References | Namespace, Name, IDs, Access Time | Platform | | |
| | Platform/Environment | | Platform | | |
| | Pipeline Steps | Step Number | Implementation | | |
| | | Name | | | |
| | | Description | | | |
| | | Version * | | | |
| | | Prerequisites * | Platform | | |
| | | Input List | Data (Input Data) | | |
| | | Output List | | Consistency | |
| Execution | Script | | Implementation | | |
| | Script driver | | Platform | | |
| | Algorithmic tools and Software Prerequisites | Name | Platform | | |
| | | Version | | | |
| | | URI | Platform | | |
| | | Access Time | | | |
| | | SHA1 Checksum | | | |
| | External Data Endpoints | Name | | | |
| | | URI | Platform | | |
| | Environment Variables | | | | |
| IO | Input Subdomain | Filename * | Data (Input Data) | | |
| | | URI | | | |
| | | Access Time * | | | |
| | | SHA1 Checksum * | | | |
| | Output Subdomain | Filename * | | Consistency | |
| | | Media Type | | | |
| | | URI | | | |
| | | Access Time * | | | |
| | | SHA1 Checksum * | | | |
| Parametric * | | Parameter | Data (Parameter) | | |
| | | Value | | | |
| | | Step | | | |
| Error * | Algorithmic Subdomain | Inclusion Rule | | Consistency | |
| | | Reference | | | |
| | Empirical Error | Inclusion Rule | | | |
| | | Reference | | | |
| | | Definitions | | | |
| Extension * | | Free Text | | | |

Table 3: A mapping summary for both frameworks: the PRIMAD Model and the BioCompute Object.

For clarity, some subfields have been excluded An asterisk (*) indicates optional domains and fields.

**PRIMAD.Consistency:** *output subdomain* of the BCO.IO and *output list* of the BCO.Description are compulsory. Although the BCO.Error is optional; it is used to define the tolerance range for input to ensure that its result is accepted. Neither has clear and direct corresponding dimensions in the PRIMAD model. We mapped *output subdomain* of the BCO.IO, the *output list* of the BCO.Description and BCO.Error onto PRIMAD.Consistency. In the use case, nine values of the error range with their subfield's description appear in BCO.Error. There are four output files in the BCO.IO on the HIVE platform and 16 files for the pipeline in the BCO.Description (twelve files for step 1 and four files for step 2).

**PRIMAD.Transparency:** We did not distinguish between the recommended and compulsory attributes for reproducibility. The PRIMAD model was extended to cover the Transparency Dimension to include all the required elements of the experiment's life cycle and guarantee reproducibility. For PRIMAD.Transparency: in the general mapping, we include all BCO compulsory fields within compulsory domains. Moreover, some compulsory fields within compulsory domains couldn't be directly mapped to a specific dimension. As a result, they are assigned to PRIMAD.Transparency. With the use case mapping, we added the BCO.Error and BCO.Parametric. The BCO.Extension was not mapped because it was empty for this use case. It is optional and has extra information that follows another schema for a pipeline in addition to the main BCO schema (IEEE 2791-2020). This information can be related to any part of a workflow. Correspondingly, this domain can be mapped onto all PRIMAD dimensions except for the Research Objective. However, we don't have the bioinformatic knowledge to distinguish between the compulsory BCO domain fields and required fields for reproducibility.

## 4. Discussion

The mapping between PRIMAD dimensions and the BCO domains showed that the reproducibility perspective underlying one framework could be incorporated into the other and also identified improvements to each framework to better guarantee reproducibility. The outcomes of this study are twofold:

- The importance of applying conceptual frameworks in practice as we identify strengths and weaknesses in the PRIMAD model ;
- A demonstration of the utility of a systematic examination of standards and models that are intended to advance adequate reproducibility reporting, with recommendations formulated for improving the BCO standard.

In the following sections, we will discuss key observations related to reproducibility within both frameworks, 4.1 and 4.2. We will then present our recommendations to address the reproducibility gaps identified in both frameworks through this study, section 4.3 for PRIMAD and 4.4 for BCO.

## 4.1 Entanglements and Redundancies in the Frameworks' Models

A prominent observation was "entanglements" and redundancies in reporting in the Dimension/Domains of both models. PRIMAD Platform, Method and Implementation Dimensions are entangled with each other (Chap et al., 2019). Entanglement refers to the interdependent relations among some dimensions, where specifications or changes in one dimension can significantly impact the others. Although a separation between different properties helps support comprehensive workflow reporting (Goble et al., 2019). In practice, the lines are blurred, especially in the experiments of a given workflow type (Haendel et al., 2016). Consequently, a script updated in PRIMAD.Implementation, may involve the effects impact PRIMAD.Method and PRIMAD.Platform. In BCO_022531/2.2, using HIVE as an execution platform necessitates adjustments to the BCO's study methods. PRIMAD.Method in principle maps to BCO.Usability, a free text to describe the BCO as a whole, and BCO.Description, a structured description for the BCO that is not meant for computation. However, for this use case, PRIMAD.Method maps to BCO.Usability only. The BCO.Description is HIVE-tailored as the pipeline's two steps were: "HIVE-Hexagon" and "Sequence Profiling Engine", where "HIVE" is the workflow's platform and holds input and output data files. BCO.Execution is platform-dependent, as might be expected, as it is mapped to the PRIMAD.Implementation.

Similarly, there are seemingly redundancies across some BCO fields, such as input and output files appearing in both BCO.Description and BCO.IO. However, this distinction is necessary to identify files used at the execution level including intermediates across steps (BCO.Description), versus the subset of "main" files indicated as meaningful for the whole computational workflow at a study level (BCO.IO).

## 4.2 PRIMAD Abstraction versus BCO Specifics

The PRIMAD model is proposed as a conceptual approach to framing reproducibility in scientific computational experiments. The BCO standard is considered an implemented approach with more detailed and tailored for NGS pipelines, which is a step towards guaranteeing these pipelines' reproducibility. The mapping between the PRIMAD model and the BCO framework reflects the demand for both to have the other's perspective/view of reproducibility. The PRIMAD model dimensions are conceptual; the dimension name and its definition; there is no checklist for each dimension. In contrast, each domain in the BCO framework has a group of defined fields. In practice, the abstractions in the PRIMAD model need depth and detail tailored to the requirements of each scientific discipline.

Conversely, the specifications in the BCO framework would benefit from an extra level of abstraction. Reporting an experiment in a conceptual manner eases portability, reimplementation and long-term reproducibility. Experiment software is subject to transformation as time progresses (Freire and Chirigati, 2018). The BCO offers a certain level of abstraction when detailing the pipeline within BCO.Usability and BCO.Description,. In BCO.Usability, researchers are advised to include a free text to report the need, method, results and how the results can be used/interpreted. While BCO.Description has structured fields to capture information in a standardized format for FDA submission. It includes pipeline steps' aspects (*step_number, name, description, version, prerequisite, input_list, output_list*), external references(*xref*), compatible platforms(*platform*) and *keywords*. In practice, when we consider BCO_022531/2.2, BCO.Usability reported a method outlines while BCO.Description reported pipeline steps that are platform dependent, *pipeline_steps.name:*"HIVE-Hexagon". Despite that, there is no mapping between the method outlines (BCO.Usability) and pipeline steps (BCO.Description) to encourage porting the experiment to another platform. A comprehensive and systematic separation of the platform implementation from the pipeline abstraction supports pipeline reproducibility using different platforms (experiment portability) or different implementations (method reinterpretation).

## 4.3 Extending the PRIMAD Model for Reproducible Pipelines in Practice

In addition to Transparency and Consistency as non-direct dimensions for PRIMAD (Freire et al., 2016, pp. 128–132), Freire and Chirigati (2018) suggested two extra non-direct dimensions for the PRIMAD model: Coverage and Longevity. **Coverage** means the portion of experiments that can be reproduced (Freire et al., 2012). Concerning some approaches, full reporting does not always guarantee reproducibility because this principle is affected by different issues. These issues can be related to costs, complicated or customised preparations of data or software, data sensitivity and copyright and/or licencing issues. **Longevity** refers to the ability to reproduce experiments long after their creation. This dimension's recommendation is attributed to the nature of software, which may change from time to time or become obsolete. As we will discuss next, capturing the times at which software is created, updated, and accessed points to longevity (Grayson et al., 2023).

The power of the PRIMAD model lies in the fact that it is a reproducibility framework for computational experiments conducted in all scientific disciplines. It has been adopted in studies to articulate certain aspects of reproducibility, including information retrieval (IR) experiments (Breuer et al., 2019; Breuer et al., 2022), neural information retrieval (Ferro et al., 2016), music information retrieval(MIR) (Knees et al., 2022), high-performance computing (HPC) (Chapp et al., 2020), natural sciences (Gryk and Ludäscher, 2017), scientific experiments' provenance (Samuel and König-Ries, 2022) and astronomy (Chapp et al., 2019). Nevertheless, using the model with its general perspective and its dimensions poses challenges to researchers in reporting experiment details.

Ensuring reproducibility requires avoiding a one-size-fits-all approach (Stewart et al., 2021). The PRIMAD model's abstraction makes it applicable to a wide range of computational experiments, but it is expected to be extended and customised to provide tailored checklists in practice to each scientific discipline (Chap et al., 2019) (Freire et al., 2016, pp. 128–132). The systematic capture of provenance (Freire and Chirigati, 2018) for each dimension facilitates reproducibility, which in turn satisfies the requirements for depth in each dimension, as identified in this investigation.

A scientific workflow is a crucial component of computational science. It refers to a series of tasks, processes, or steps that are directed towards achieving a computational goal or result. Providing a FAIR framework with which to develop a computational workflow is anticipated to improve the reproducibility and transparency of the workflow life cycle (Goble et al., 2019). Our detailed examination of a genomic data workflow analysis using the BCO framework highlighted the necessity of expanded depth and detail in each dimension of the PRIMAD model to reproduce these computational pipelines. Inspired by the BCO framework, we propose ten cross-cutting aspects for the PRIMAD Dimensions: time, additional resources, data categorisations, functionality, versions, human roles, licences, fault tolerance, reviews, and metadata schema. Table [4] identifies the related BCO subfields as examples of each aspect.

| Time | | Additional Resources | |
|---|---|---|---|
| BCO Fields | Domains | BCO Fields | Domains |
| Obsolescence Time | Provenance | External References | Description |
| Embargo Period | | Fault Tolerance | |
| Created Time | | BCO Fields | Domains |
| Modification Time | | Empirical Error Values | Error |
| Review Date | | Algorithm Error Values | Error |
| Input/output Subdomain Access Time | I/O | Metadata Schemas | |
| Input/output List Access Time | Description | BCO Fields | Domains |
| External Reference Access Time | | BCO Schema version | Top-Level Fields |
| Data Categorizations | | Additional Schema | Extension |
| BCO Fields | Domains | Version | |
| Input/output Subdomains | I/O | BCO Fields | Domains |
| Input/output Lists | Description | BCO Version | Provenance |
| Parameters | Parametric | Software Prerequisites Version | Execution |
| Human Roles | | Pipeline Step Version | Description |
| BCO Fields | Domains | Licence | |
| Contributor | Provenance | BCO fields | Domains |
| Reviewer | | BCO licence | Provenance |
| Review | | | |
| BCO Fields | Domains | BCO fields | Domains |
| Review Status | Provenance | Review Date | Provenance |
| Reviewer Comment | | Reviewer | |

| Operationality | | | |
|---|---|---|---|
| BCO Fields | Domains | BCO Fields | Domains |
| Platform/Environment | Description | Modification Time | Provenance |
| Pipeline Step's Prerequisite | | BCO Version | |
| SHA Checksum | Execution | Obsolescence Time | |
| Environment Variable | | Embargo Period | |
| External Data Endpoints | | BCO licence | |
| Software Prerequisites | | SHA Checksum | I/O |
| | | ETag | Top-Level Fields |

Table 4: Proposed aspects for the PRIMAD model and their documentation through the BCO domains' fields.

In addition to the current exploration, Willis (2020) attempted to present a list of elements required for reproducible computational research artifacts. This list can serve as a checklist for preparing computational experiments for reproducibility. We recommend incorporating our proposed aspects into various dimensions, as detailed in Table [5]. In the following discussion, each aspect and its corresponding PRIMAD dimensions are discussed in detail, <u>including how the aspect contributes to ensuring reproducibility, the involved PRIMAD dimensions, and the related BCO fields.</u>

| Proposed Aspects / PRIMAD Dimensions | Data Categorizations | Versions | Time | Additional Resources | Fault Tolerance | Licence | Functionality | Human Roles | Review | Metadata Schemas |
|---|---|---|---|---|---|---|---|---|---|---|
| Research Objective | | ✓ | | | | | | | ✓ | ✓ |
| Method | | | | | | | | | | |
| Platform | | ✓ | ✓ | ✓ | ✓ | | ✓ | | | ✓ |
| Implementation | | ✓ | ✓ | ✓ | ✓ | ✓ | ✓ | | | ✓ |
| Actor | | | | | | | | ✓ | | ✓ |
| Data | ✓ | ✓ | ✓ | ✓ | ✓ | ✓ | ✓ | | | ✓ |

Table 5: The proposed depth is in the form of aspects for each PRIMAD Model's dimension.

**Data Categorisation:** The Data Dimension requires reporting all the data used or produced during a study. Simultaneously, data should be classified into input, intermediate and output, each with its own rules for treatment in an experiment (Chap et al., 2019). Data in the BCO are categorized into input, output, and parameter, see Table[4].

**Version:** Awareness of the versions of the data repository or library, script driver and software tools that represent the resources of a computational experiment is a significant issue, as are the versions of data and script files. Additionally, it is important to maintain awareness of the versions for metadata schemas, or any standards utilized throughout the computational study. A lack of such awareness can impede the reproduction of a study. Furthermore, as in Table[5,] we propose a version aspect for the Research Objective Dimension that reflects the study/experiment version. The versions of a study itself, if it has more than one, should maintain provenance and inheritance; as *version* in BCO.Provenance, see Table [4].

**Time:** The time aspect adds timestamps for various events in the lifecycle. Access to resources is limited in time, so researchers should consider running studies for a suitable period to enable them to use all facilities while they are available (forbidden as *embargo* in BCO.Provenance), operational and active (not shut down or expired as *obsolete_after* in BCO.Provenance), see Table [4]. Tracking time after a study is completed is equally important, and tracking time after modification creates a new version. Additionally, tracking access time can identify the activities and conditions related to research resources.

**Additional Resources:** While a computational experiment is rerun, different resources are accessed and/or used. These resources can come in the form of a platform, database, repository, schema, ontology, and/or script library, referenced using *xref* in BCO.Description (External References) Table [4]. Identifying the resource type, its associated experiment elements with their identifiers or citations (data, script, actor, etc.), its requirements for environment configuration and its prerequisites for access and use should be indicated in experiment reporting.

**Fault Tolerance:** To reproduce a study, variations in input and intermediate data may be allowed and are essential to demonstrating accepted tolerance levels. This approach also aligns with acceptable variations for environmental variables. The level of output variations should be reported to rule on the success of reproducibility. These variations serve to declare wriggle room that moderates fixity. Additionally, this variation can be used to optimize the algorithm within *empirical_error* and *algorithmic_error* in BCO.Error, Table [4]. PRIMAD.Consistency, as an extra dimension in the model, expresses fault tolerance for the result in a non-directional manner.

**Licences:** Reporting a licence for reusing code or script in computational experiments is recommended (Willis, 2020). Understanding the copyright status of the study and its data is also necessary before pursuing reproducibility. Anonymization techniques (Abdulmajeed & Lee, 2021) could be applied for data for studies where data are confidential or sensitive. In Table[4], *license* in BCO.Provenance provides a license for the BCO as a whole Research Object, rather than per individual data files and scripts.

**Functionality:** Ensuring the data files and scripts/codes are functional before reproducing an experiment is essential. The execution resources/infrastructure, such as servers, platforms, libraries, repositories, and software tools, should also be functional, as discussed in the "additional resources" aspect. Before utilizing a file, data or script, one should verify its validity, confirm that it is appropriately updated, and ensure compatibility with the execution environment prerequisites. Moreover, it is necessary to assess its accessibility and confirm that all necessary permissions are provided.

The BCO adopts the functionality aspect through many fields, as shown in Table[4]. For validity, the (*etag*) includes a string generated by a Secure Hash Algorithm, a SHA-256 hash function, to protect the object from alterations without proper validation. Also, this algorithm is optional when applied to input and output files. Besides this algorithm, including the BCO expiration date (*obsolete_after*) reflects the study's intended validity period. BCO provides many fields that ensure the requirements of the execution environment, such as the *prerequisite* of the pipeline step, *xrf* (external references), *environment_variables*, *software_prerequisites* and *external_data_points* (network requirements) Recording modification dates of study data, along with their versions and details of changes, enhances the study's maintainability. In addition, the *platform* field presents the alternative's compatible platforms. Usage permission is employed in (*embargo*) and (*licence*) fields.

For the computational experiment, the constraints that could affect the file and resource's functionality differ from one discipline to another. These constraints could relate to time passage (platform, software or file format no longer used), specific authority (files are sensitive or confidential), property rights (property software or platform), complex configuration, etc. Consequently, some fields that address the proposed aspect of 'functionality' also encompass other proposed aspects, such as time, version, licensing, and additional resources. This explains why some BCO fields are duplicated for functionality in Table[4]. We recommend reporting the functionality of the files and the different resources with respect to the nature of the scientific discipline. This can be achieved by applying continuous testing (Hettne et al., 2012) that monitors validity, maintainability, accessibility and usage permission.

**Human Roles:** In computational experiments, even with the automation of most or all experiment steps, humans can play vital roles in executing certain procedures. They create, observe, make decisions, review, examine, reason and so on. However, such roles are under-reported or difficult to document in current reporting conventions. Accordingly, for the PRIMAD model, additional attributes are needed to document the duties that humans assume in an experiment (Chap et al., 2019). These roles can be described using the Contributor Roles Taxonomy (CRediT), which identifies 14 roles for research contributors (Brand et al., 2015), and several more detailed taxonomies have been proposed for research outputs and software contributions (Peroni et al., 2018; Alliez et al., 2019). The BCO standard provides a list of contribution roles based on the PAV ontology (Ciccarese et al., 2013): { *authoredBy, contributedBy, createdAt, createdBy, createdWith, curatedBy, derivedFrom, importedBy, importedFrom, providedBy, retrievedBy, retrievedFrom, sourceAccessedBy* }. At the same time, including the research contact details could mitigate any inquiry about reproducing the study.

**Reviews:** Acquiring knowledge of review outcomes prior to reproducing a specific study may enhance its reproducibility (Willis, 2020). This can help identify obstacles within the review process—factors that may complicate reproducibility. The review process is reported in BCO.Provenance including: {*status, reviewer_comment, date, reviewer*}. For the PRIMAD model, we recommend making reporting the details of a review optional because the model can be used to evaluate the study itself.

**Metadata Schemas:** Metadata facilitates experiment reporting. Leipzig et al. (2021) argue that metadata for a computational experiment can turn the reproducibility crisis into an opportunity. The BCO framework features a compulsory main field for the schema that describes a given version of this BCO specification: https://w3id.org/ieee/ieee-2791-schema/2791object.json. Naturally, a change to the metadata schema itself can cause semantic or syntactic differences across the other fields, and so the schema version should be explicit, as shown by BCO. Additionally, the BCO framework can link another schema in BCO.Extension for additional structured information, for instance, fhir extension for embedding metadata following the Fast Health Interoperability Resources (FHIR) standard (Ayas et al., 2021). Standard metadata types are descriptive, administrative, structured metadata and markup languages (NISO, 2017). Metadata can be domain-specific or cross-domain elements. They can be human- or machine-readable, or both. A BCO specification is structured human- and machine-readable metadata, while CRediT is descriptive human-readable metadata. Schema.org is a notable metadata used to engage semantics in web pages. CWL is a workflow and tool metadata for sharing across different pipeline platforms (Crusoe et al., 2022). PRIMAD model dimensions serve as metadata that describe the key components of the computational experiments. The depth of each dimension can be further detailed using our proposed aspects or alongside a domain-specific metadata standard tailored to the discipline's requirements.

## 4.4 BCO Reconfiguration

The investigation through the PRIMAD model and BCO standard motivated the idea of reconfiguring some of the BCO fields to emphasise reproducibility. This reconfiguring enables the clearer and more abstract viewing of workflow details by researchers outside the NGS discipline and maintains detailed experiment reporting for long-term reproducibility by discipline experts. Refer back to Table[2], which outlines the main BCO domains and their primary functionalities. In the following discussion, we offer recommendations for some specific BCO domains.

**Data and Method Consistency across Domains**

For the computational experiment, the method and data description should be consistent with the implementation facilitating the experiment's portability (port) and reinterpretation (re-code), see Table[1]. Consistency in the data description is achieved by declaring the input, output, and intermediate files and their relation to the method and implementation. Additionally, the methodology consistency ensures that the method and associated data can be directly mapped to the corresponding implementation. This can be observed in the BCO.Usability and BCO.Description through the BCO schema specification. However, as discussed in section 4.2, the use case (BCO_022531/2.2) did not adopt the methodology consistency between BCO.Usability and BCO.Description.

In the BCO, Data are mentioned in the BCO.IO, BCO.Description and BCO.Parametric domains, and these include data files and paramateres that have been used or obtained as results, while methods are mentioned in BCO.Usability and BCO.Description domains. To ensure consistency in the description of data and methods across BCO domains, we recommend the following steps:

- Merge BCO.Usability and BCO.Description into one domain that describes the workflow independently from the platform. The merged domain will include BCO.Usability text that describes the objectives, method outlines, results, and result interpretation structurally in the form of definite fields; and input/output list, platform/environment (compatible platforms within BCO), keywords and external references (XRF) fields from BCO.Description. The proposed merged domain will represent a conceptual view of the workflow in a systematic view (definite fields). We could suggest naming it with Conceptual Domain.
- Expand BCO.Execution to include the *pipeline_steps* fields that existed previously in BCO.Description and associated with step script, except input/output list and platform/environment fields.
- Retain BCO.IO and BCO.Parametric as domains for the input, output and parameter, as they are platform dependent.

In the Conceptual domain, the proposed domain, we suggest structuring its fields as follows:

- **Method:** The experiment methodology should be platform-independent and enriched with sufficient information to apply it to a different platform (port) or implementation (recode). This can be achieved by linking the method outlines to the pipeline steps and the input/output data files, which are platform-independent.

- **External Reference:** In addition to its list of the databases and/or ontology IDs that are cross-referenced in the BCO, it could include the names of scientific theoretical methods used in the workflow associated with their published references. Typing each reference will help explain their role in the BCO.
- **Keywords:** Provide the Keyword field with a list that accurately captures the core content of the experiment. The field can consist of terms relating to aspects such as objectives, methods, platform names, database and ontology names, contributors' names, and their affiliations.
- **Meaningful Names for Files:** Emphasise assigning meaningful names to input and output files in order to be understandable for the external researcher (Trisovic et al., 2022).

**Enhance the Provenance Domain with Reproduction Studies and Publications**

Documenting previous reproducibility studies and their outcomes is crucial, as it sustains time and effort for future reproducibility efforts or review processes. Also, it highlights any necessary adjustments to the experiment and ensures the documentation of scientific efforts for researchers. Additionally, including the resulting scientific publications or any type of dissemination of the experiment will link the experiment to the associated scientific merit. Therefore, we suggest extensions to the BCO.Provenance as follows:

- **BCO Reproduction Provenance:** We recommend documenting the provenance of the previous reproducibility processes performed in a study. This can be achieved by adding fields to capture the details of the reproducibility process made over the BCO framework, including the date, result, obstacles, coverage (Freire et al., 2012), and a new BCO's reference if it exists.
- **BCO-Related Disseminations:** We suggest including any related scientific publications identifiers, such as DOIs, to enable new studies to reference how the works are documented in these publications in addition to the BCOs themselves. . In addition to the publications, it could include any other type of study activity, such as talks, presentations, or posters, in the form of digital object identifiers in any medium or platform through which this experiment has been published or shared.

**Permission for Resources Usage across Domains**

To enhance the functionality of the various resources used in the experiment, we recommend to include a field for each referenced BCO resource (such as data, scripts, library and platforms) indicating whether it is open access, requires authorization, or is not authorized. Justifications should be provided, such as the presence of sensitive data or proprietary software, along with details on how to obtain authorization if possible.

# Conclusion

This study pursues two primary objectives: assess the reproducibility of using the BioCompute Object as a framework for reporting the NGS pipeline; and evaluate the suitability of the PRIMAD model as a reproducibility framework for computational workflows, specifically in genomic data. These Objectives are achieved by mapping both frameworks onto a published BCO use case (BCO_022531/2.2) (King et al., 2022).

The result confirms the PRIMAD model's applicability to computational experiments, as it lays down broad guidelines for the approaches used in different disciplines. On the other hand, our examination of the model proposed extensions to support reproducing the computational workflow (pipeline). The extensions identified ten aspects to expand the model dimensions. These aspects are inspired by the BCO framework, which are time, additional resources, data categorisations (Chap et al. (2019)), functionality, versions, human roles (Chap et al. (2019)), licences, fault tolerance, reviews, and metadata schemas.

Unequivocally, our findings support the BioCompute Object Framework's promotion of reproducibility. The BCO demonstrated its strength in its tailored reproducibility framework for the NGS pipeline reporting. However, the meticulous mapping of both frameworks onto a published BCO use case (BCO_022531/2.2) revealed some omissions regarding BCO reproducibility. Therefore, we recommend the necessity of including a systematic conceptual reporting of the computational workflow in the form of definite fields in the schema. This reporting should be platform-independent and aligned with the workflow's data and implementation. It will support Longevity (Freire and Chirigati, 2018) and aid its portability to another execution environment if the experiment software/platform is no longer usable or accessible. In addition, we make some recommendations to the BCO Provenance Domain for reporting reproduction trials and related publications.

This study identifies key attributes for achieving reproducibility in computational workflow in bioinformatics, such as through checklists and replication kits. Our findings underscore the importance of implementing tools specifically designed to aid researchers in their reporting tasks. Moreover, we advocate for further investigation into applying the PRIMAD model with aspects of reproducibility across different scientific disciplines. An approach based on this common conceptual model will enhance the reproducibility of computational science.

# Acknowledgements

Many thanks to Charles Hadley King, Jonathan Keeney and Raja Mazumder for giving us the authorization to access and view the use case files (BCO_022531/2.2) at the HIVE platform.

# Data Records

The supplementary data for this paper are available on Zenodo (**DOI:** 10.5281/zenodo.14317922) as an RO-Crate, offering a structured reporting approach for the study and its associated assets.

38. Freire, J. and Chirigati, F. (2018). Provenance and the Different Flavors of Computational Reproducibility. IEEE Data Engineering Bulletin, 41(1), pp. 15-26, IEEE. https://web.archive.org/web/20220119190028id_/http://sites.computer.org/debull/A18mar/p15.pdf

39. Breuer, T. (2020). Reproducible online search experiments. In J. M. Jose, E. Yilmaz, J. Magalhães, P. Castells, N. Ferro, M. J. Silva, & F. Martins (Eds.), *Advances in information retrieval: 42nd european conference on IR research, ECIR 2020, lisbon, portugal, april 14–17, 2020, proceedings, part II*, Lecture notes in computer science (Vol. 12036, pp. 597–601). Cham: Springer International Publishing. DOI: 10.1007/978-3-030-45442-5_77

40. Breuer, T., Keller, J., & Schaer, P. (2022). ir_metadata: An Extensible Metadata Schema for IR Experiments. *Proceedings of the 45th International ACM SIGIR Conference on Research and Development in Information Retrieval* (pp. 3078–3089). Presented at the SIGIR '22: The 45th International ACM SIGIR Conference on Research and Development in Information Retrieval, New York, NY, USA: ACM. DOI: 10.1145/3477495.3531738

41. Ferro, N., Fuhr, N., Järvelin, K., Kando, N., Lippold, M., & Zobel, J. (2016). Increasing reproducibility in IR. *ACM SIGIR Forum*, *50*(1), 68–82. DOI: 10.1145/2964797.2964808

42. Knees, P., Ferwerda, B., Rauber, A., Strumbelj, S., Resch, A., Tomandl, L., ... & Dizdar, R. (2022, December). A Reproducibility Study on User-centric MIR Research and Why it is Important. In *Proceedings of the 23rd International Society for Music Information Retrieval Conference*. https://humrec.github.io/publication/knees-ismir-2022-b/knees-ismir-2022-b.pdf

43. Chapp, D., Stodden, V., & Taufer, M. (2020).Building a vision for reproducibility in the cyberinfrastructure ecosystem: leveraging community efforts. Supercomputing Frontiers and Innovations, 7(1). DOI: 10.14529/jsfi200106

44. Gryk, M. R., & Ludäscher, B. (2017). Workflows and provenance: toward information science solutions for the natural sciences. Library trends, 65(4), 555–562. DOI: 10.1353/lib.2017.0018
30